\documentclass[aps,prb,amsmath,amssymb,superscriptaddress,reprint]{revtex4-1}

\usepackage{graphicx}
\usepackage{dcolumn}
\usepackage{bm}

\begin{document}

\bibliographystyle{unsrt}

\title{Giant ferroelectric polarization and electric reversal of strong spontaneous magnetization in multiferroic Bi$_2$FeMoO$_6$}

\author{Peng Chen}
\affiliation{Beijing National Laboratory for Condensed Matter Physics, Institute of Physics, Chinese Academy of Sciences, Beijing 100190, China}
\author{Bang-Gui Liu}%
\email{bgliu@iphy.ac.cn} \affiliation{Beijing National Laboratory for Condensed Matter Physics, Institute of Physics, Chinese Academy of Sciences, Beijing 100190, China}
\affiliation{School of Physical Sciences, University of Chinese Academy of Sciences, Beijing 100190, China}


\date{\today}

\begin{abstract}
BiFeO$_3$ is the most famous multiferroic material, but it has no strong
spontaneous magnetization due to its antiferromagnetism. Here we show that giant ferroelectric polarization and strong spontaneous magnetization can be both realized in double
perovskite Bi$_2$FeMoO$_6$ with R3 (\#146) space group based on BiFeO$_3$. Our first-principles
phonon spectra establishes that this multiferroic R3 phase is stable. Our systematic
calculations show that it is a spin-polarized semiconductor with gap reaching
to 0.54 eV and has a strong ferroelectric polarization of 85$\mu$C/cm$^2$. This
ferroelctricity is comparable with that of BiFeO$_3$, but here
obtained is a strong ferrimagnetism with net magnetic moment of 2$\mu_B$ per formula unit and Curie temperature of 650 K.
Both ferroelectric polarization and magnetic easy axis are shown to be in
pseudocubic [111] orientation. Our further analysis shows that the macroscopic spontaneous magnetization can be deterministically reversed through a three-step path by external electric field. Therefore, we believe that this Bi$_2$FeMoO$_6$ material can be used to design new multifunctional materials and achieve high-performance devices.
\end{abstract}


\pacs{75.85.+t, 75.30.-m, 77.80.-e, 75.10.-b}
\keywords{multiferroics, magnetoelectric, $BiFeO_3$}

\maketitle


\section{Introduction}

One of the most significant advantages of magnetoelectric (ME)
multiferroics is that the ME coupling\cite{RME,MEMu1} can be used to manipulate magnetic
properties by external electric fields, or to control electric properties by external magnetic field\cite{MDevice2}. However, one of challenges is to
find promising multiferroic materials that have both large magnetization and
enormous ferroelectric polarization beyond room temperature. So far, the
perovskite BiFeO$_3$ (BFO) is the only such single-phase multiferroic material
\cite{BFOTc1,BFOTc2,BFOTc3,BFOTc4,MDevice1,RevBFO1,RevBFO2,ExpBFO,DFTBFO}.
Both bulk and thin film BFO samples can exhibit strong ferroelectricity ($P$ $\simeq$ 100$\mu$C/cm$^2$) with
high Curie temperature 1100 K \cite{DFTBFO} and strong
antiferromagnetism with magnetic Neel temperature 650 K
\cite{BFOMag1,BFOMag2,BFOMag3,BFOMag4}. The
Dzyaloshinskii-Moriya (DM)\cite{BFODMI1,BFODMI2} interaction can
induce in it a weak ferromagnetism with a remanent magnetization of
$0.02 \mu_B/$Fe\cite{WFerr,t1}. It has been revealed theoretically
and experimentally that BiFeO$_3$-based systems can display novel
coupling phenomenon between spontaneous electric polarization in the
pseudocubic [111] direction and the weak ferromagnetism
perpendicular to the [111] direction\cite{t1,switch2}, which makes it
possible to control the weak ferromagnetism through the electric
polarization switching by an external electric field.

A strong macroscopic magnetization above room temperature is highly
desirable for energy-efficient spintronic
devices\cite{MDevice1,MDevice2}. Here, we show giant ferroelectric polarization and electric reversal of strong macroscopic spontaneous magnetization in multiferroic Bi$_2$FeMoO$_6$ (BFMO) phase with R3 space group. Its high-temperature ferrimagnetism have been proposed \cite{lsd} and its multiferroic phase (with a small ferroelectric polarization and a magnetic Curie temperature beyond room temperature) has been realized experimentally \cite{new1}. Our
first-principles calculations reveal that it is a semiconductor with
gap reaching to 0.54 eV, and has ferroelectric polarization of
85$\mu$C/cm$^2$ and strong spontaneous ferrimagnetism with net magnetic moment
of 2$\mu_B$ per formula unit. Considering its ferroelectricity is similar to that of the BiFeO$_3$, we believe that its ferroelectric polarization can persist beyond room temperature as its
macroscopic spontaneous magnetization does.
Furthermore, our study shows that its magnetic easy axis is
collinear with its ferroelectric polarization vector in the pseudocubic [111]
directions, and its spontaneous magnetization can be deterministically reversed
through three switching steps of its ferroelectric polarization under
an applied electric field. More detailed results will be presented
in the following.

\section{Computational Methods}

We use projector augmented-wave
(PAW)\cite{paw} plus pseudo-potential methods within the density
functional theory\cite{DFT1,DFT2}, as implemented in the Vienna ab
initio simulation package (VASP) \cite{vasp1,vasp2}, to optimize the
crystal structures and then study the electronic structures,
ferroelectricity, and magnetic properties. The spin-polarized
generalized-gradient approximation (GGA)\cite{pbe}, to the
electronic exchange-correlation functional, is used to do structure
optimization. To make the structures compatible with different
magnetic configurations and atomic distortions, calculations were
performed using a 40-atom super-cell which can be considered as
doubling the ideal perovskite structure along the three Cartesian
directions. Plane wave basis set with a maximum kinetic energy of
500 eV and $3\times 3\times 3$ k-point mesh generated with the
Monkhorst-Pack scheme\cite{mp} were used. During optimization, all
of the structures were fully relaxed until the largest force between
the atoms become less than 1 meV/\AA{}. When we obtain very small total energy difference between two structures in terms of the pseudo-potential method, we use the full-potential method to more accurately describe the total energy difference in order to improve the total energy comparison\cite{wien2k}.

Phonon spectra are calculated with first-principles perturbation
method with norm-conserving pseudo-potentials as implemented in
package Quantum-ESPRESSO \cite{phonon}. The magnetic anisotropy
energy (MAE) due to the spin-orbit coupling is determined by the
force theorem, and calculated in terms of the total energy with
respect to polar and azimuthal angle along the [111] direction.
Switching paths were determined using the `Nudged Elastic Band'
(NEB)\cite{NEB} method which can give the most energetically
favourable intermediate configuration between the initial and final
states. We use the modified Becke-Johnson exchange potential to
accurately calculate semiconductor gaps\cite{wien2k,mbj}. We use modern
Berry phase method\cite{berry} and take both GGA+U
\cite{ldau1,ldau2} and HSE06 \cite{hse1,hse2} schemes to calculate
electric polarization.

\section{Results and discussions}

\subsection{Optimized structure}

{\it Possible crystalline and magnetic structures.} Double
perovskite Bi$_2$FeMoO$_6$ (BFMO) can be related with the famous perovskite BiFeO$_3$ (BFO) by considering that BFMO is obtained through partially substituting Fe in BFO by Mo. In order to seek
stable phases of BFMO, we start with 12 basic space groups of Bi$_2$FeMoO$_6$ allowed
by group analysis\cite{g1,g2,g3,g4,g5}, and then we optimize the
crystal structures and calculate the total energies by considering
three ferroelectric distortions and four magnetic configurations:
one ferromagnetic (FM) and three antiferromagnetic (AFM) or
ferrimagnetic (FIM) orders of A-type, C-type, and G-type. Actually,
all possible structures related with the 12 space groups have been
considered, including the different magnetic structures and
possible ferroelectric distortions of Bi atoms in three cartesian directions.
These cover cubic, tetrahedral, orthorhombic, rhombohedral,
hexagonal, and monoclinic crystal systems. The most stable six
phases are summarized in Table \ref{tab:struct}, where we present
their space groups, Glazor notes for oxygen-octahedron tilting,
volumes, net magnetic moments, and total energies (with the lowest
energy set to zero). It is interesting and reasonable that the most stable magnetic
configuration is the FIM order of G-type for all the six structures
shown in Table \ref{tab:struct}. Our results also show that the R3
phase is semiconductive, the P$2_1/n$ and R$\bar{3}$ are half-metallic,
and the other three are normally metallic.

\begin{table}
\caption{Space groups, Glazor notes, volumes ($V$, in \AA$^3$/f.u.),
magnetic moments ($M$, in $\mu_B$/f.u.), and total energies
($\Delta$$E$, in meV/f.u.) of the lowest six of the structures.}\label{tab:struct}
\begin{ruledtabular}
\begin{tabular}{cccccc}
No. & Space group   &Glazor  &$V$  & $M$ & $\Delta E$ \\
\hline
1 & R$3$       &$a^-a^-a^-$    &133.27  & 2    &0\\
2 & P$2_1/n$   &$a^-a^-c^+$    &128.62  & 2    &75. \\
3 & R\={3}     &$a^-a^-a^-$    &131.94  & 2    &401.  \\
4 & C2/m       &$a^0b^-b^-$    &129.15  & 2.5    &364.  \\
5 & I4/m       &$a^0a^0c^-$    &127.93  & 2.3    &763.  \\
6 & Pn\={3}    &$a^+a^+a^+$    &128.89  & 2.2    &921.  \\
\end{tabular}
\end{ruledtabular}
\end{table}

\begin{figure*}[!htbp]
\includegraphics[width=13cm]{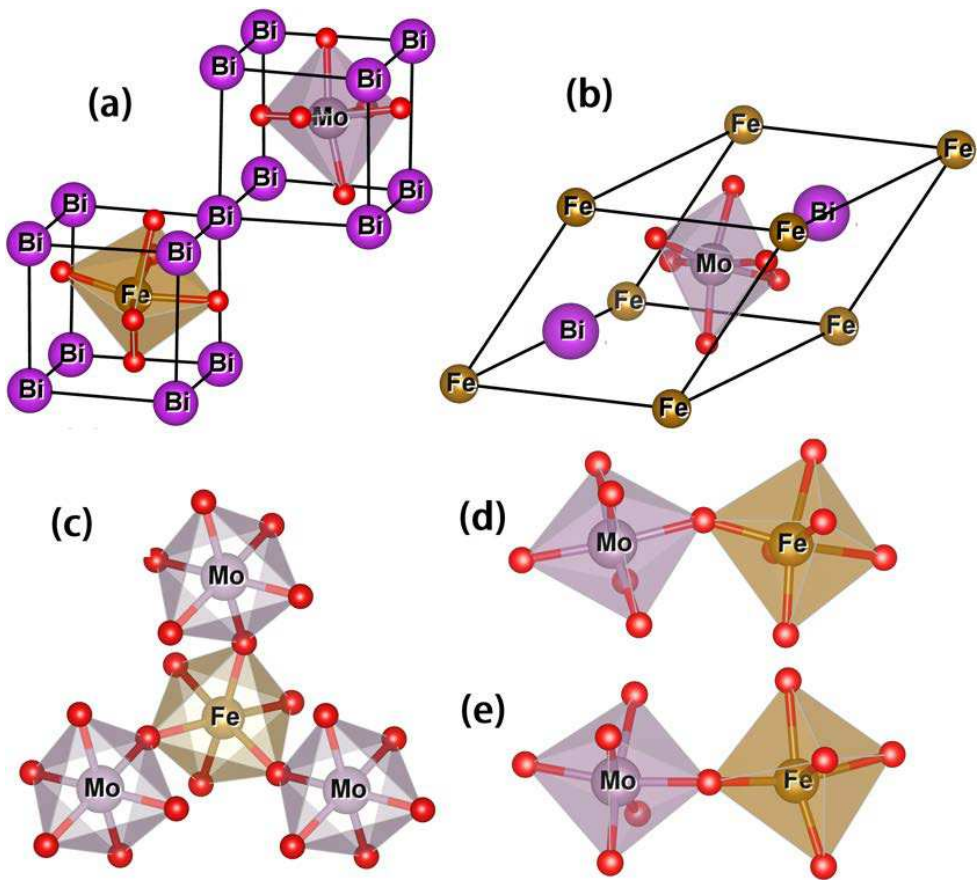}
\caption{\label{fig1} The key structural features: (a) the crystal
structure of BFMO showing the alternate occupation of Fe and Mo at
the B-sites along the [111] direction, (b) the primitive cell of
BFMO with the O octahedron of Mo shown, (c) the relative rotation
projected on the (111) plane of an O octahedron of Fe with respect
to the nearest O octahedra of Mo, and (d) and (e) the two
neighboring O octahedra of Fe and Mo projected on the plane of the
Fe-O-Mo triangle and projected on the perpendicular plane including
the Fe and Mo ions.}
\end{figure*}

{\it The R3 phase as a multiferroic material.} Our total energy
comparison reveals that the R3 is the ground state phase with the
lowest energy, and it exhibits ferroelectric property. There exists a P$2_1/n$ metastable magnetic
phase which is higher by only 75 meV per formula unit than the ground state phase.
This R3 phase can be considered to be distorted from the
R$\bar{3}$ structure, and is comparable with the ground state R3c
bulk phase in the case of BFO. Here, we focus on the R3 phase
because it has strong ferrimagnetic order, large ferroelectric
polarization, and useful magneto-electric coupling, as will be shown
in the following. We illustrate the crystal structure, local
octahedral tilting, and crystal deformation in Fig \ref{fig1}. The
O-Fe-O, O-Mo-O, and Fe-O-Mo bond angles are reduced to
163.3$^\circ$, 168.6$^\circ$, and 150.0$^\circ$, respectively, which
reflect substantial crystal deformation. We have calculated the
phonon spectra of the P$2_1/n$ and R3 phases and present that of the R3
structure in Fig. 2(a). These results show that there is not any
instability in these two cases. Because the other phases are at least 75 meV/f.u. higher than the rhombohedral R3 phase,
we believe that this R3 phase can be synthesized as epitaxial (111)
thin films on appropriate substrates, like those of
BFO\cite{TF1111,TF1112,TF1113,TF1114}.

\begin{figure*}[!htbp]
\includegraphics[width=17cm]{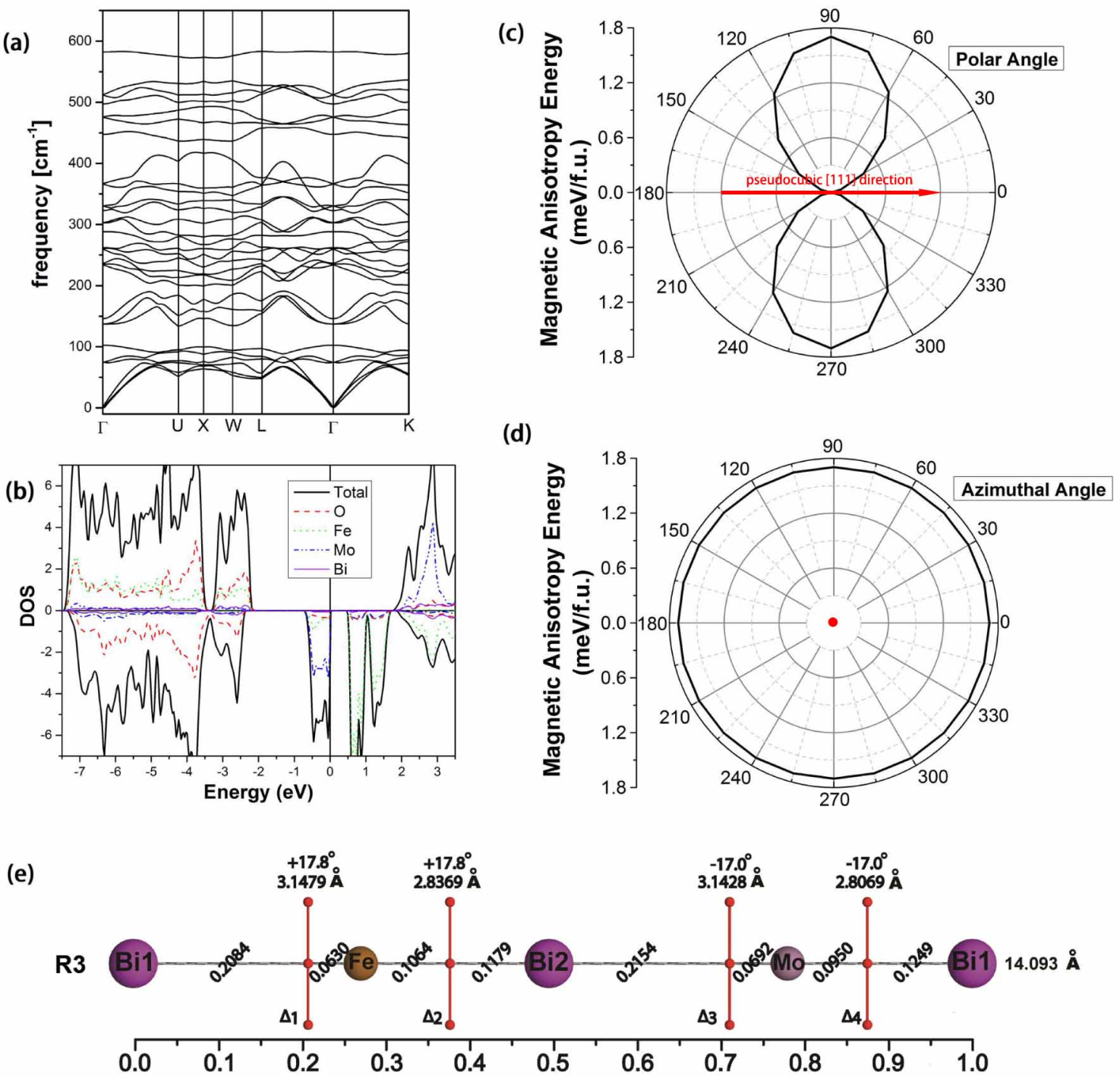}
\caption{\label{fig2} Basic properties of BFMO: (a) the electronic
density of states (DOS), (b) the phonon spectra showing no
structural instability, (c) and (d) the relative total energies in a
plane including the [111] axis and in the (111) plane, and (e) an
analysis of ion displacements. The DOS in (a), in state/eV per
primitive cell, shows a net moment 2$\mu_B$ and a semiconductor gap
0.54 eV, and the upper half shows spin-up DOS and the lower
spin-down DOS. The energies in (c,d) shows a uniaxial easy axis in
[111] and an MCA energy of 1.5 meV. The skew numbers on the
horizontal line in (e) are the relative distances between the
positive ions (Bi1, Fe, Bi2, Mo) and the centers of the nearest O
triangles ($\Delta_1$, $\Delta_2$, $\Delta_3$, $\Delta_4$), and the
numbers above the vertical lines show the relative rotational angles
around the [111] axis and the edge lengthes of the O triangles,
respectively. There is a $60^{\circ}$ rotation between $\Delta_1$
and $\Delta_2$, and between $\Delta_3$ and $\Delta_4$.}
\end{figure*}

\subsection{Electronic structure and magnetism}

Because GGA usually
underestimate semiconductor gaps, we use the modified Becke-Johnson
(mBJ) exchange functional to study the electronic structure.
Presented in Fig \ref{fig2}(b) is our mBJ calculated spin-dependent
density of states (DOS) of the optimized R3 structure of
Bi$_2$FeMoO$_6$. Our mBJ calculated gap, 0.54 eV, should be a much
better value for the true gap because mBJ has been proved to give
accurate semiconductor gaps for most of semiconductors and, more
importantly, our mBJ gap for BFO is equivalent to 2.1 eV which
is in good agreement with experimental results of 2.5
eV\cite{BFOGap2,BFOGap3}. Our analysis reveals that in the spin-up
DOS, the three filled Fe t2g bands merges into the oxygen p bands
and the two filled Fe eg bands are separated from them, and the
lowest empty bands, mainly from Mo 4d states, are above 1.75 eV. It
can be seen that the semiconductor gap, 0.54 eV, is formed
between the three filled Mo 4d bands and the three empty Fe 3d ones
in the spin-down channel. Totally, we have five Fe d electrons in
the spin-up channel and three Mo d electrons in the spin-down
channel, and thus the magnetic moment per formula unit is equivalent
to 2$\mu_B$, which is consistent with the calculated data. We also
use other exchange-correlation schemes to calculate the
semiconductor gap and thereby make sure that the
semiconductor gap is true for the R3 phase of Bi$_2$FeMoO$_6$. On the other
hand, we also calculate spin exchange energies and Curie
temperatures for the R3 phase, confirming the high magnetic Curie
temperature of 650 K\cite{lsd}.

\subsection{Excellent multiferroics}

We have studied the effect of the spin-orbit
coupling for the R3 phase. We present the calculated
magnetocrystalline anisotropy energy (MAE) as functions of the polar
and azimuthal angles in Fig. \ref{fig2} (c) and (d). Here, the polar
and azimuthal angles are defined with respect to the pseudo[111]
direction. In contrast to the magnetic easy plane for
BFO\cite{t1,switch1}, BFMO has a magnetic easy axis and its
magnetization tends to align along the pseudocubic [111] axis.
Actually, the MAE can be well described by a simple expression,
$E_{\bf MA}=1.5\sin^2\theta$ in meV, where the polar angle $\theta$
takes values from 0 to $\pi$. On the other hand, we have analyze the
ionic positions of the R3 phase projected in the [111] direction and
show the result in Fig \ref{fig2} (e). We observe clear cation
displacements with respect to their oxygen environment, which are
like those in the BFO which exhibits strong ferroelectricity.
Furthermore, we quantitatively investigate the ferroelectricity of
BFMO through modern DFT-based calculations with both
GGA+U\cite{ldau1,ldau2} and HSE06\cite{hse1,hse2} methods. For the
GGA+U scheme, we use one set of parameter (U,J)= (2.5eV,0.5eV) for Mo and three sets for Fe: (U,J)=(2eV,0.8eV), (4eV,0.8eV), and (8eV,0.8eV). Our calculations give a large
spontaneous ferroelectric polarization $85. \mu $C/cm$^2$ in the
[111] direction, which is nearly independent of calculational
schemes and parameters. In addition to the high magnetic
Curie temperature, we expect a high
ferroelectric Curie temperature for this ferroelectricity because it should share the similar mechanism of ferroelectricity with BFO. Therefore, BFMO should be an excellent
multiferroic material.

\begin{figure*}[!htbp]
\includegraphics[width=14cm]{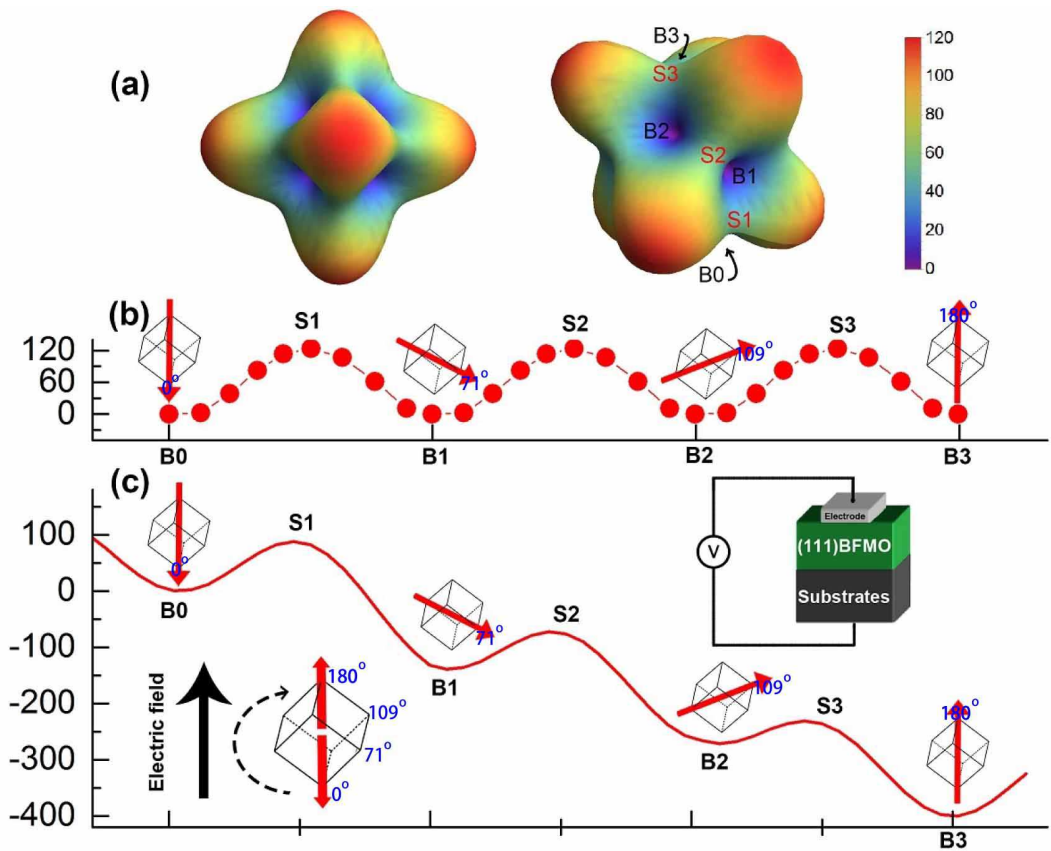}
\caption{\label{fig3} Reversal of the uniaxial magnetization driven
by electric three-step switching of polarization. (a) collinear
reversal of polarization through $-P \rightarrow 0 \rightarrow +P$,
with energy barriers of 340 meV/f.u.; (b) rotational reversal of
polarization through three steps $0^{\circ} \rightarrow 71^{\circ}
\hookrightarrow 109^{\circ} \rightarrow 180^{\circ}$, with energy
barriers of 120 meV/f.u.; and (c) the three-step reversal of the
electric polarization and the uniaxial magnetization with an
electric field applied in the $180^{\circ}$ direction. The polar
angles accrued at B1, B2, and B3 are $71^{\circ}$, $109^{\circ}$,
and $180^{\circ}$, respectively. Both of the vectors actually rotate
by $71^{\circ}$ for each step, with the second step contributing
only $38^{\circ}$ to the polar angle.}
\end{figure*}

\subsection{Electric reversal of magnetization}

{\it Cubic anisotropy and reversal paths of ferroelectric polarization.} Actually, there
are eight equivalent [111] directions, which implies that the
ferroelectric anisotropy is cubic\cite{switch1,switch2}. We can describe the ferroelectric
crystalline anisotropy by the following expression,
\begin{equation}
E_f=K[1/3-(\alpha_1^2\alpha_2^2+\alpha_2^2\alpha_3^2+\alpha_3^2\alpha_1^2)],
\end{equation}
where ($\alpha_1$,$\alpha_2$,$\alpha_3$) denotes the
directional cosine of the ferroelectric polarization vector and our calculation gives $K=1.44$ eV. The ferroelectric polarization anisotropy is plotted in Fig. 3(a), where there are eight minima or bottom points (B0, B1, B2, B3 etc.) and twelve saddle points (S1, S2, S3 etc). Assuming usual collinear
polarization reversal over the R$\bar{3}$ structure as an
intermediate one, $-P \rightarrow 0 \rightarrow +P$, our DFT
calculation gives an energy barrier 340 meV per formula unit. In addition, we have found a three-step path for
the polarization reversal: $0^{\circ} \rightarrow 71^{\circ}
\hookrightarrow 109^{\circ} \rightarrow 180^{\circ}$, where the three nonzero angles describe the polar angles of B1, B2, and B3 with respect to B0. Along this
path, the polarization vector actually changes the equivalent angle
of 71$^{\circ}$ three times, but the second step is not in the polar
angle direction, contributing to only 38$^{\circ}$ to the polar
angle, as shown in Fig. 3(b). The barrier for each of the three
steps, the energy difference between the saddle points and the bottom points, is equivalent to 120 meV per formula unit, substantially lower
than that of the collinear reversal. Therefore, this three-step reversal path is
favorable in energy.

{\it Electric three-step reversal of strong magnetization.} As we have
shown above, the magnetic [111] easy axis is collinear with the
ferroelectric polarization vector along the [111] direction. The MA
energy can be re-written as $E_{\bf MA}=c_{me}(\vec{M}\times
\vec{P})^2$, where $\vec{M}$ is magnetization vector and $c_{me}$ ($>0$)
describes the magneto-electric coupling strength. As shown in Fig. 3(c), if we apply an electric field antiparallel to the ferroelectric polarization vector, the field can make the energy curve substantially decline. For each of the energy wells, because the left barrier is substantially lower than the right barrier, the polarization vector will try to follow the field. Applying a rotational electric field, from 0 to 180$^{\circ}$, can make the polarization vector follow the three-step path for the polarization reversal: $0^{\circ} \rightarrow 71^{\circ}
\hookrightarrow 109^{\circ} \rightarrow 180^{\circ}$. Because of $E_{\bf MA}$, the magnetization tends to stay in the same direction as the ferroelectric polarization vector and will finally reverse, as long as the rotation of the electric field is slow enough.

\section{Conclusion}

In conclusion, we have investigated double
perovskite Bi$_2$FeMoO$_6$ through systematic DFT calculations and
structural analyses. Our careful GGA optimization reveals that the
Bi$_2$FeMoO$_6$ in the R3 structure is a ferrimagnetic
semiconductor and the net moment reaches to 2$\mu_B$ per formula unit
because the magnetic moments of Fe and Mo cannot be canceled. The
semiconductor gap has been corrected with mBJ potential, reaching to
0.54 eV. Our DFT-based polarization calculations directly shows
that the Bi$_2$FeMoO$_6$ has strong ferroelectric polarization 85$\mu$C/cm$^2$ in the
pseudocubic [111] direction. Therefore, we believe that the strong ferroelectricity in
the Bi$_2$FeMoO$_6$ can have high Curie temperature far beyond
room temperature, because its mechanism should be similar to that of BFO. With the spin-orbit effect into account, our
calculation show that the magnetic easy axis is also in the [111]
axis. Because the ferroelectric polarization and magnetic easy axis are both in
pseudocubic [111] orientation, we show that the strong macroscopic spontaneous magnetization can be deterministically reversed through a three-step path by external electric field. When realized, the high-quality Bi$_2$FeMoO$_6$ material, like BiFeO$_3$, can be useful for designing new multifunctional materials and achieve high-performance devices in the future.

\begin{acknowledgments}
This work is supported by the Nature Science Foundation of China (No. 11574366), by the Strategic Priority Research Program of the Chinese Academy of Sciences (Grant No.XDB07000000), and by the Department of Science and Technology of China (Grant No. 2016YFA0300701).
\end{acknowledgments}


\begin{thebibliography}{99}

\bibitem{RME} M. Fiebig, J. Phys. D: Appl. Phys. \textbf{38}, R123 (2005).

\bibitem{MEMu1} W. Eerenstein, N. D. Mathur and J. F. Scott, Nature \textbf{442}, 759 (2006).

\bibitem{MDevice2} S. Fusil, V. Garcia, A. Barthelemy and M. Bibes, Annu. Rev. Mater. Res. \textbf{44}, 91 (2014).

\bibitem{BFOTc1} D. M. Evans, A. Schilling, A. Kumar, D. Sanchez, N. Ortega, M. Arredondo, R. S. Katiyar, J. M. Gregg and J. F. Scott, Nat. Commun. \textbf{4}, 1534 (2013).

\bibitem{BFOTc2} A. A. Belik, S. Iikubo, K. Kodama, N. Igawa, S.-I. Shamoto, S. Niitaka, M. Azuma, Y. Shimakawa, M. Takano, F. Izumi and E. Takayama-Muromachi, Chem. Mater. \textbf{18}, 798 (2006).

\bibitem{BFOTc3} M.-R. Li, U. Adem, S. R. C. McMitchell, Z. Xu, C. I. Thomas, J. E. Warren, D. V. Giap, H. Niu, X. Wan, R. G. Palgrave, F. Schiffmann, B. Slater, T. L. Burnett, M. G. Cain, A. M. Abakumovi, G. Tendelooi, M. F. Thomas, M. J. Rosseinsky and John B. Claridge, J. Am. Chem. Soc. \textbf{134}, 3737 (2012).

\bibitem{BFOTc4} W. Wang, J. Zhao, W. Wang, Z. Gai, N. Balke, M. Chi, H. N. Lee, W. Tian, L. Zhu, X. Cheng, D. J. Keavney, J. Yi, T. Z. Ward, P. C. Snijders, H. M. Christen, W. Wu, J. Shen and X. Xu, Phys. Rev. Lett. \textbf{110}, 237601 (2013).

\bibitem{MDevice1} D. Sando, A. Barthelemy and M. Bibes, J. Phys.: Condens. Matter \textbf{26}, 473201 (2014).

\bibitem{RevBFO1} J.-G. Park, M. D. Le, J. Jeong and S. Lee, J. Phys.: Condens. Matter \textbf{26}, 433202 (2014).

\bibitem{RevBFO2} G. Catalan and J. F. Scott, Adv. Mater. \textbf{21}, 2463 (2009).

\bibitem{ExpBFO} J. Wang, J. B. Neaton, H. Zheng, V. Nagarajan, S. B. Ogale, B. Liu,
D. Viehland, V. Vaithyanathan, D. G. Schlom, U. V.  Waghmare, N. A. Spaldin, K. M. Rabe, M. Wuttig and R. Ramesh,
Science \textbf{299}, 1719 (2003).

\bibitem{DFTBFO} J. B. Neaton, C. Ederer, U. V. Waghmare, N. A. Spaldin and K. M. Rabe, Phys. Rev. B \textbf{71}, 014113 1 (2005).

\bibitem{BFOMag4} J.-M. Moreau, C. Michel, R. Gerson and W. J. James, J. Phys. Chem. Solids \textbf{32}, 1315 (1971).

\bibitem{BFOMag2} Y. E. Roginskaya and Y. Tomashpolskii, Zh. Eksp. Teor. Fiz \textbf{50}, 69 (1966).

\bibitem{BFOMag3} W. Kaczmarek, M. Polomska and Z. Pajak, Phys. Lett. A \textbf{47}, 227 (1974).

\bibitem{BFOMag1} S. Kiselev, G. Zhdanov and R. Ozerov, Dokl. Akad. Nauk SSSR \textbf{145}, 1255 (1962).

\bibitem{BFODMI1} I. A. Sergienko and E. Dagotto, Phys. Rev. B \textbf{73}, 094434 (2006).

\bibitem{BFODMI2} T. Arima, A. Tokunaga, T. Goto, H. Kimura, Y. Noda and Y. Tokura, Phys. Rev. Lett. \textbf{96}, 097202 (2006).

\bibitem{WFerr} H. Bea, M. Bibes, S. Petit, J. Kreisel and A. Barthelemy, Philos. Mag. Lett.  \textbf{87}, 165 (2007).

\bibitem{t1} C. Ederer and N. A. Spaldin, Phys. Rev. B \textbf{71}, 060401 (2005).


\bibitem{switch2} J. T. Heron, J. L. Bosse, Q. He, Y. Gao, M. Trassin, L. Ye, J. D. Clarkson, C. Wang, J. Liu, S. Salahuddin, D. C. Ralph, D. G. Schlom, J. Iniguez, B. D. Huey and R. Ramesh, Nature \textbf{516}, 370 (2014).


\bibitem{lsd} S.-D. Li, P. Chen and B.-G. Liu, AIP Advances \textbf{3}, 012107 (2013).

\bibitem{new1} S. Ravi and C. Senthilkumar, Mater. Express \textbf{5}, 68 (2015).


\bibitem{paw} P. E. Blochl, Phys. Rev. B \textbf{50}, 17953 (1994).

\bibitem{DFT1} P. Hohenberg and W. Kohn, Phys. Rev. \textbf{136}, B864 (1964).

\bibitem{DFT2} W. Kohn and L. J. Sham, Phys. Rev. \textbf{140}, A1133 (1965).

\bibitem{vasp1} G. Kresse and J. Hafner, Phys. Rev. B \textbf{47}, 558 (1993).

\bibitem{vasp2} G. Kresse and J. Furthmuller, Phys. Rev. B \textbf{54}, 11169 (1996).

\bibitem{pbe} J. P. Perdew, K. Burke and M. Ernzerhof, Phys. Rev. Lett. \textbf{77}, 3865 (1996).

\bibitem{mp} H. J. Monkhorst and J. D. Pack, Phys. Rev. B \textbf{13}, 5188 (1976).

\bibitem{wien2k} P. Blaha, K. Schwarz, G. K. H. Madsen, D. Kvasnicka, and J. Luitz, WIEN2k, An Augmented Plane Wave + Local Orbitals Program for Calculating Crystal Properties, Karlheinz Schwarz Technische Universitat Wien, Austria, 2001, ISBN 3-9501031-1-2.

\bibitem{phonon} P. Giannozzi, S. Baroni, N. Bonini, M. Calandra, R. Car, C. Cavazzoni, D. Ceresoli, G. L. Chiarotti, M. Cococcioni and I. Dabo, J. Phys.: Condens. Matter \textbf{21}, 395502 (2009).

\bibitem{NEB} G. Mills, H. Jonsson and G. K. Schenter, Surf. Sci. \textbf{324}, 305 (1995).

\bibitem{mbj} F. Tran and P. Blaha, Phys. Rev. Lett. \textbf{102}, 226401 (2009).

\bibitem{berry} R. D. King-Smith and D. Vanderbilt, Phys. Rev. B \textbf{47}, 1651 (1993).

\bibitem{ldau1} V. I. Anisimov, J. Zaanen and O. K. Andersen, Phys. Rev. B \textbf{44}, 943 (1991).

\bibitem{ldau2} V. I. Anisimov, F. Aryasetiawan and A. Lichtenstein, J. Phys.: Condens. Matter \textbf{9}, 767 (1997).

\bibitem{hse1} J. Heyd, G. E. Scuseria and M. Ernzerhof, J. Chem. Phys. \textbf{118}, 8207 (2003).

\bibitem{hse2} J. Heyd, J. E. Peralta, G. E. Scuseria and R. L. Martin, J. Chem. Phys. \textbf{123}, 174101 (2005).

\bibitem{g1} A. M. Glazer, Acta Crystallogr. B \textbf{28}, 3384 (1972).

\bibitem{g2} A. M. Glazer, Acta Crystallogr. A \textbf{31}, 756 (1975).

\bibitem{g3} P. M. Woodward, Acta Crystallogr. B \textbf{53}, 44 (1997).

\bibitem{g4} C. J. Howard and H. Stokes, Acta Crystallogr. B \textbf{54}, 782 (1998).

\bibitem{g5} C. J. Howard, B. J. Kennedy and P. M. Woodward, Acta Crystallogr. B \textbf{59}, 463 (2003).


\bibitem{TF1111} J. Li, J. Wang, M. Wuttig, R. Ramesh, N. Wang, B. Ruette, A. Pyatakov, A. Zvezdin and D. Viehland, Appl. Phys. Lett. \textbf{84}, 5261 (2004).

\bibitem{TF1112} F. Bai, J. Wang, M. Wuttig, J. Li, N. Wang, A. Pyatakov, A. Zvezdin, L. Cross and D. Viehland, Appl. Phys. Lett. \textbf{86}, 032511 (2005).

\bibitem{TF1113} M. K. Singh, H. M. Jang, S. Ryu and M.-H. Jo, Appl. Phys. Lett. \textbf{88}, 042907 (2006).

\bibitem{TF1114} H. Bea, M. Bibes, X. Zhu, S. Fusil, K. Bouzehouane, S. Petit, J. Kreisel and A. Barthelemy, Appl. Phys. Lett. \textbf{93}, 072901 (2008).


\bibitem{BFOGap2} T. Kanai, S.-I. Ohkoshi and K. Hashimoto, J. Phys. Chem. Solids \textbf{64}, 391 (2003).

\bibitem{BFOGap3} S. J. Clark and J. Robertson, Appl. Phys. Lett. \textbf{90}, 132903 (2007).



\bibitem{switch1} M. Lezaic and N. A. Spaldin, Phys. Rev. B \textbf{83}, 024410 (2011).



\end{thebibliography}
\end{document}